\documentclass[journal=aamick,manuscript=article]{achemso}

\usepackage{chemformula} % Formula subscripts using \ch{}
\usepackage[T1]{fontenc} % Use modern font encodings
\usepackage[version=3]{mhchem}
\usepackage{amsmath}
\usepackage{fdsymbol} 

\author{Mitchell Albert}
\affiliation{Department of Materials Science and Engineering, McMaster University, 1280 Main Street West, Hamilton, Ontario L8S 4L8, Canada}
\author{Amanda Clifford}
\affiliation{Department of Materials Science and Engineering, McMaster University, 1280 Main Street West, Hamilton, Ontario L8S 4L8, Canada}
\author{Igor Zhitomirsky}
\affiliation{Department of Materials Science and Engineering, McMaster University, 1280 Main Street West, Hamilton, Ontario L8S 4L8, Canada}
\email{zhitom@mcmaster.ca}
\author{Oleg Rubel}
\affiliation{Department of Materials Science and Engineering, McMaster University, 1280 Main Street West, Hamilton, Ontario L8S 4L8, Canada}
\phone{+1 905 5259140 ext. 24295}
\email{rubelo@mcmaster.ca}

\title[Adsorption of maleic acid\ldots]
  {Adsorption of maleic acid monomer on the surface of hydroxyapatite and \ce{TiO2}: a pathway toward biomaterial composites}


\begin{document}

\begin{tocentry}

\includegraphics{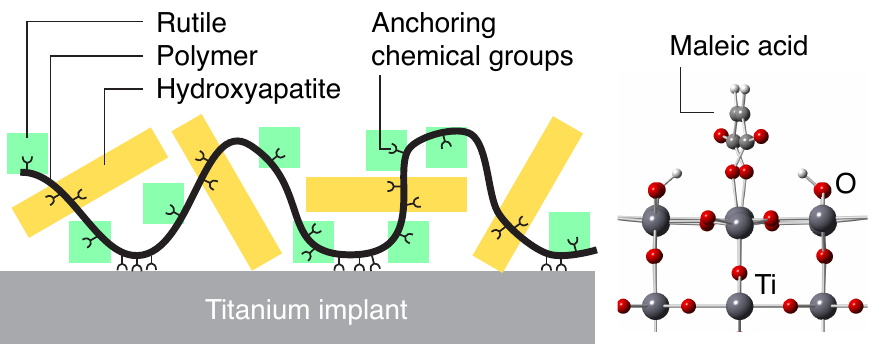}

\end{tocentry}

\begin{abstract}
Poly(styrene-alt-maleic acid) adsorption on hydroxyapatite and \ce{TiO2} (rutile) was studied using experimental techniques and complemented by \textit{ab initio} simulations of adsorption of  a maleic acid segment as a subunit of the copolymer. \textit{Ab initio} calculations suggest that the maleic acid segment forms a strong covalent bonding to the \ce{TiO2} and hydroxyapatite surfaces. If compared to vacuum, the presence of a solvent significantly reduces the adsorption strength as the polarity of the solvent increases. The results of first-principle calculations are confirmed by the experimental measurements. We found that adsorbed poly(styrene-alt-maleic acid) allowed efficient dispersion of rutile and formation of films by the electrophoretic deposition. Moreover, rutile can be co-dispersed and co-deposited with hydroxyapatite to form composite films. The coatings showed an enhanced corrosion protection of metallic implants in simulated body fluid solutions, which opens new avenues for the synthesis, dispersion, and colloidal processing of advanced composite materials for biomedical applications.
\end{abstract}

\section{Introduction}
Investigations of adsorption of organic molecules at the surface of nanomaterials and their colloidal behaviour\cite{Xu_JMC_17_2007} allowed for the development of novel strategies for the surface modification, dispersion, and advanced synthesis of nanoparticles. Studies of mussel adsorption on inorganic surfaces\cite{Lee_S_318_2007,Lee_ARMR_41_2011} provide important chemical and physical insights into development of covalent anchoring mechanisms. It was found that the strong mussel adhesion involves protein macromolecules that contain a catecholic amino acid \mbox{L-3,4-dihydroxyphenylalanine} (L-DOPA). The adhesion mechanism of mussels is attributed to the complexation, or bridging bidentate bonding between metal atoms on material surfaces and hydroxyl groups of catechol.\cite{Hidber_JECS_17_1997} These studies have generated interest in the investigation of catecholates and inspired the development of advanced adhesives,\cite{Lee_ARMR_41_2011,Lee_N_448_2007} dispersants,\cite{Ata_RA_4_2014} liquid-liquid extraction agents\cite{Clifford_ML_201_2017,Chen_JCIS_499_2017,Poon_CI_43_2017,Wallar_CSA_500_2016} containing anchoring catechol groups for various applications (see Ref.~\citenum{Ye_CSR_40_2011} and references therein).

The success in the applications of chelating molecules from the catechol family has driven investigations of natural aliphatic compounds with carboxyl groups, which can provide strong chelating or bridging polydentate bonding to inorganic materials. Of particular interest are fumaric and maleic acid (MA) isomers, containing \textit{trans} and \textit{cis} carboxyl groups, respectively. These acids showed strong adsorption on various oxides.\cite{Dobson_SAMBS_55_1999} The \textit{cis} conformation of maleic acid provides an ideal orientation for coordination \textit{via} both carboxylate groups in a strong tetradentate interaction.\cite{Dobson_SAMBS_55_1999,Lenhart_JCIS_345_2010} It is in this regard that many dispersant molecules described in the literature provide a relatively weak monodentate bonding to the particle surface.\cite{Ata_RA_4_2014,Pujari_ACIE_53_2014,Gaponik_NL_2_2002,Gao_S_1_2005,Paul_CSA_482_2015} Clearly, tetradentate interactions of maleic acid with oxide surfaces offer advantages for surface modification of materials.

\begin{figure}%[h]
\centering
  \includegraphics[width=0.3\columnwidth]{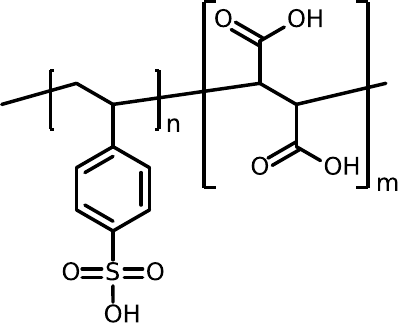}
  \caption{Styrene-maleic acid copolymer.}
  \label{Fig:1}
\end{figure}

Significantly stronger interactions with inorganic surfaces can be expected using maleic acid polymers or copolymers. The individual maleic acid monomers of such polymers can provide multiple chemical bonds with substrates. Poly(maleic acid) showed a strong adsorption on alumina particles and allowed for their efficient dispersion.\cite{Mohanty_JACS_1_2013} Poly(acrylic acid-co-maleic acid) exhibited a strong affinity to \ce{BaTiO3} and clay minerals.\cite{Blockhaus_JCIS_186_1997,Zhao_CI_30_2004} The adsorption properties of poly(styrene-alt-maleic acid) (PSMA-h) copolymer (Fig.~\ref{Fig:1}) have been utilized for the synthesis of inorganic particles with different morphologies\cite{Lai_JPCC_113_2009,Yu_JSSC_177_2004,Yu_JSSC_179_2006} and unusual superstructures.\cite{Xu_AM_20_2008} PSMA-h is a biocompatible polymer, which is currently under investigations for many biomedical applications, such as drug delivery,\cite{Henry_B_7_2006,Dalela_AAMI_7_2015,Angelova_CSA_452_2014} biosensors,\cite{Baghayeri_SAB_188_2013} antimicrobial materials\cite{Samoilova_JPCB_113_2009} and implants.\cite{Saez-Martinez_RFP_94_2015,Hongkachern_AMR_93-94_2010} PSMA-h demonstrated a strong affinity to hydroxyapatite (HA), bioglass, and bioceramics and allowed their electrophoretic deposition (EPD) and co-deposition with proteins.\cite{Clifford_CSA_516_2017} Colloidal techniques, such as EPD, have a high potential in the development of advanced films, coatings, scaffolds and devices for biomedical applications. Therefore, the investigation of PSMA-h adsorption on bioceramics opens new avenues for the synthesis, dispersion, and colloidal processing of advanced materials for biomedical applications.

HA is an important material for biomedical implant applications, because its chemical composition is similar to that of natural bone.\cite{Boccaccini_JRSI_7_2010} Rutile is also known as a bioactive material, which promotes HA biomineralization.\cite{Cai_RA_6_2016,Lindahl_JMSMM_21_2010,Lu_JMSMM_21_2010} Compared to the anatase phase, the rutile phase of \ce{TiO2} has many advantages for implant applications, such as phase stability, chemical stability and improved corrosion protection.\cite{Kasemanankul_MCP_117_2009} The addition of rutile to the HA to form a composite coating offers many benefits, such as enhanced chemical stability, improved corrosion protection of implants, bioactivity, improved mechanical properties, enhanced osteoblast adhesion and cell growth.\cite{Xiao_SCT_200_2006,Ulasevich_RA_6_2016,Jaworski_JTST_19_2010,Hannora_JAC_658_2016,Sarao_MMTA_43_2012}  Previous investigations\cite{Clifford_CSA_516_2017} showed that HA coatings can be deposited by EPD using PSMA-h as a charging and dispersing agent, which strongly adsorbed on the HA surface.

The goal of this investigation was to probe the PSMA-h adsorption on HA and \ce{TiO2} (rutile) using experimental techniques as well as to perform \textit{ab initio} simulations of adsorption of a maleic acid segment as a subunit of the copolymer. We identified the most stable \ce{TiO2} and HA surfaces and their reconstructions in the presence of a solvent using first-principle atomistic modelling. We examined molecular mechanisms and adsorption strength of MA on different surfaces of \ce{TiO2} and HA. Our simulations indicate that two carboxylate functional groups of MA form strong covalent bonds at the surfaces of HA and \ce{TiO2}. In these calculations MA was treated as a segment of a polymer chain rather than a monomer. Our studies demonstrate that a solvent  significantly alters the surface energy and adsorption characteristics of molecules. Experimental measurements are performed to test validity of the theory. The measurements include the use of PSMA-h as a dispersion agent for rutile and formation of films by EPD. We examined the deposition yield, film morphology, and corrosion resistance. The possibility of co-dispersing and co-depositing the rutile with HA to form composite films is investigated for biomedical applications.

%%%%%%%%%%%%%%%%%%%%%%%%%%%%%%%%%%%%%%%%%%%%%%%%%%%%%%%%%%%%%%%%%%%%%%%%%%%%%%%%%%%%%%%%%%%
\section{Methods}

\subsection{Computational}

The first-principle electronic structure calculations have been performed in the framework of the density functional theory (DFT)\cite{Kohn_PR_140_1965} using Perdew-Burke-Ernzerhof generalized gradient approximation (GGA-PBE) for the exchange-correlation functional.\cite{Perdew_PRL_77_1996} The Vienna \textit{ab initio} simulation program (VASP) and projector augmented-wave potentials\cite{Blochl_PRB_50_1994,Kresse_PRB_59_1999,Kresse_PRB_54_1996} were used. Full self-consistent structural optimization was performed for the bulk \ce{TiO2} (rutile) and HA. The structure was considered optimized when the magnitude of Hellmann-Feynman forces acting on atoms dropped below 10~meV/{\AA} and components of the stress tensor did not exceed 1 kBar. The Brillouin zone was sampled using $3\times3\times4$ and $2\times2\times4$ Monkhorst-Pack grid\cite{Monkhorst_PRB_13_1976} for \ce{TiO2} and HA primitive cells, respectively. The cutoff energy for a plane wave expansion was set at 500~eV, which is 25\% higher than the value recommended in the pseudopotential file for oxygen. The higher cutoff energy was essential for obtaining accurate structural parameters, which are summarized in Table~\ref{Tab:1}. The theoretical lattice parameters are within 2\% error of the experimental values that is typical for GGA-PBE. An attempt was made to include the on-site Coulomb interaction for Ti $d$-electrons in the framework of \citet{Dudarev_PRB_57_1998} using the effective Hubbard energy of $U=2$~eV.\cite{Hu_JPCC_115_2011} However, this correction resulted in a greater deviation between theoretical and experimental lattice parameters and thus was abandoned.

\begin{table}
\scriptsize
  \caption{\ Bulk structure properties (lattice parameters and fractional coordinates of atoms) of \ce{TiO2} and HA. Theoretical results are obtained using DFT. The neutron diffraction experimental data for \ce{TiO2} and HA are taken from Refs.\cite{Howard_ACB_47_1991,Kay_N_204_1964}.}
  \label{Tab:1}
  \begin{tabular}{l l l l l l l}
    \hline
    Compound & \multicolumn{3}{c}{Calculated} & \multicolumn{3}{c}{Experiment}\\
     & $a=b$ (\AA) & $c$ (\AA) & $u$ & $a=b$ (\AA) & $c$ (\AA) & $u$ \\
    \hline
    \ce{TiO2} (rutile) & 4.664 & 2.966 & O (0.3046, 0.3046, 0) & 4.594 & 2.959 & O (0.3048, 0.3048, 0)\\
    \hline
    Hydroxyapatite & 9.549 & 6.934 & Ca$_\text{I}$ (1/3, 2/3, 0.0016) & 9.432 & 6.881 & Ca$_\text{I}$ (1/3, 2/3, 0.0014)\\
    & & & Ca$_\text{II}$ (0.2501, 0.9986, 0.2513) & & & Ca$_\text{II}$ (0.2466, 0.9931, 1/4)\\
    & & & P (0.3988, 0.3682, 0.2524) & & & P (0.3982, 0.3682, 1/4)\\
    & & & O$_\text{I}$ (0.3301, 0.4850, 0.2538) & & & O$_\text{I}$ (0.3283, 0.4846, 1/4)\\
    & & & O$_\text{II}$ (0.5877, 0.4638, 0.2470) & & & O$_\text{II}$ (0.5876, 0.4652, 1/4)\\
    & & & O$_\text{III}$ (0.3379, 0.2549, 0.0737) & & & O$_\text{III}$ (0.3433, 0.2579, 0.0705)\\
    \hline
  \end{tabular}
\end{table}

\begin{figure}%[t]%[h]
\centering
  \includegraphics{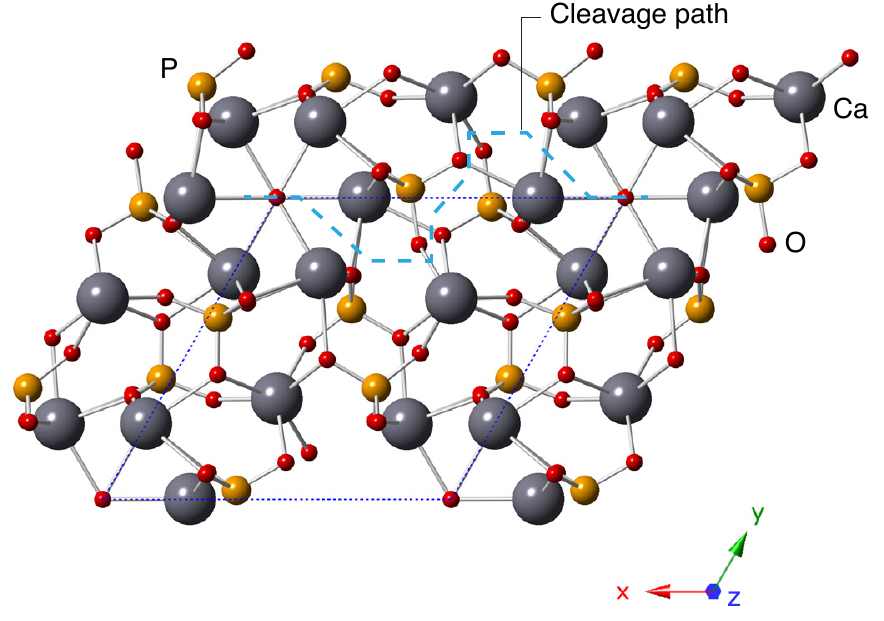}
  \caption{HA low-energy $(01\overline{1}0)$ stoichiometric surface is created by cleaving the slab along the dashed line. The construction maintains the integrity of \ce{PO4} tetrahedra.}
  \label{Fig:2}
\end{figure}

\ce{TiO2} $(110)$, HA $(0001)$ and $(01\overline{1}0)$ surfaces are obtained on the basis of the optimized bulk structures. The 20~{\AA} spacing between periodical images of slabs in direction perpendicular to the surface ensures no spurious interactions. The \ce{TiO2} $(110)$ slab is represented by four monolayers of Ti-atoms with a well-established surface termination that consists of alternating rows of 5- and 6-fold-coordinated Ti sites running along the [001] direction.\cite{Ramamoorthy_PRB_49_1994,Bullard_L_22_2006} The HA $(0001)$ and $(01\overline{1}0)$ slabs are constructed using 9 and 7 monolayers of Ca-atoms, respectively. There are some ambiguities about HA $(01\overline{1}0)$ surface and its reconstruction.\cite{Zhu_CPL_396_2004,Almora-Barrios_L_26_2010,Chiatti_TCA_135_2016} Several scenarios for a stoichiometric cleavage surface were explored. The lowest surface energy corresponds to the cleavage plane selected such that \ce{PO4} tetrahedra remain preserved (Fig.~\ref{Fig:2}). It is this surface reconstruction that is later used for studying the adsorption of molecules. Atoms of the slab representing the surface were allowed to relax except for three monolayers in the centre of the slab that were constrained to the bulk atomic positions. The presence of a solid-liquid interface is modelled using an implicit solvation model implemented in VASPsol\cite{Mathew_JCP_140_2014} and values of the static dielectric constant for a water-ethanol mixture tabulated in Ref.~\citenum{Wohlfarth2008}.

\begin{figure}%[t]%[h]
\centering
  \includegraphics{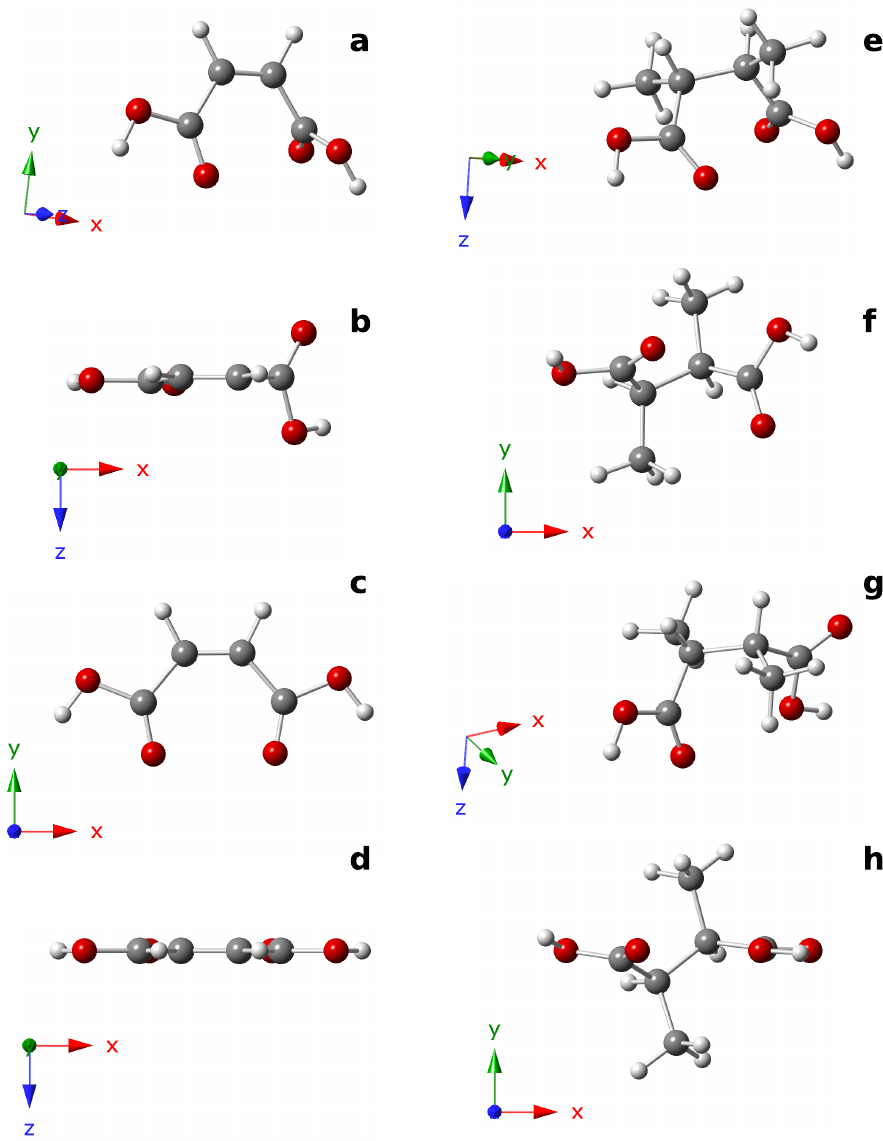}
  \caption{The structure of maleic acid as a monomer (a,b) 3D conformer, (c,d) 2D flattened conformer, and as a part of the polymer chain (e,f) 3D conformer, (g,h) 2D flattened carboxyl groups. The deformation energy associated with changing the confirmation 3D$\rightarrow$2D  is approximately 0.15~eV.}
  \label{Fig:3}
\end{figure}

The MA molecular structure (Fig.~\ref{Fig:3}a,b) is obtained by relaxing all degrees of freedom in a simulation box of the size $15\times15\times15$~{\AA}$^3$ to prevent spurious interactions between periodical images. However, this molecule is not chemically equivalent to the MA being a segment of the copolymer. To model the MA segment, two hydrogen-terminated carbon atoms were added to the MA molecule such that they mimic a polymer chain (Fig.~\ref{Fig:3}e,f). As a result, the $sp^2$ double bond between two carbon atoms in MA transforms into an $sp^3$ single bond in the MA segment. It is assumed that the individual MA segment can represent an MA residue as a part of copolymer and can serve as a basis for studying its adsorption at inorganic surfaces. This choice is driven by the necessity to make our model computationally feasible at the first-principle level. At the same time, we admit that this approach omits some details of the copolymer confirmation that can limit exposure of MA residues to the solvent/surface and possible interactions between phenyl group and the surfaces.

The adsorption of MA monomer and MA segment is modelled as an inner-sphere surface complex. The molecules were initially positioned at the surface in such a way that the distance and arrangement between the metal ion and oxygen atoms of the ligands resembles that in the bulk. The structural optimization is performed until Hellmann-Feynman forces acting on atoms dropped below 20~meV/{\AA}. Crystallographic information files (CIF) with atomic structures used in calculations can be accessed through the Cambridge crystallographic data centre (CCDC deposition numbers 1582972$-$1582992).

\subsection{Experimental}

\begin{figure}%[t]%[h]
\centering
  \includegraphics{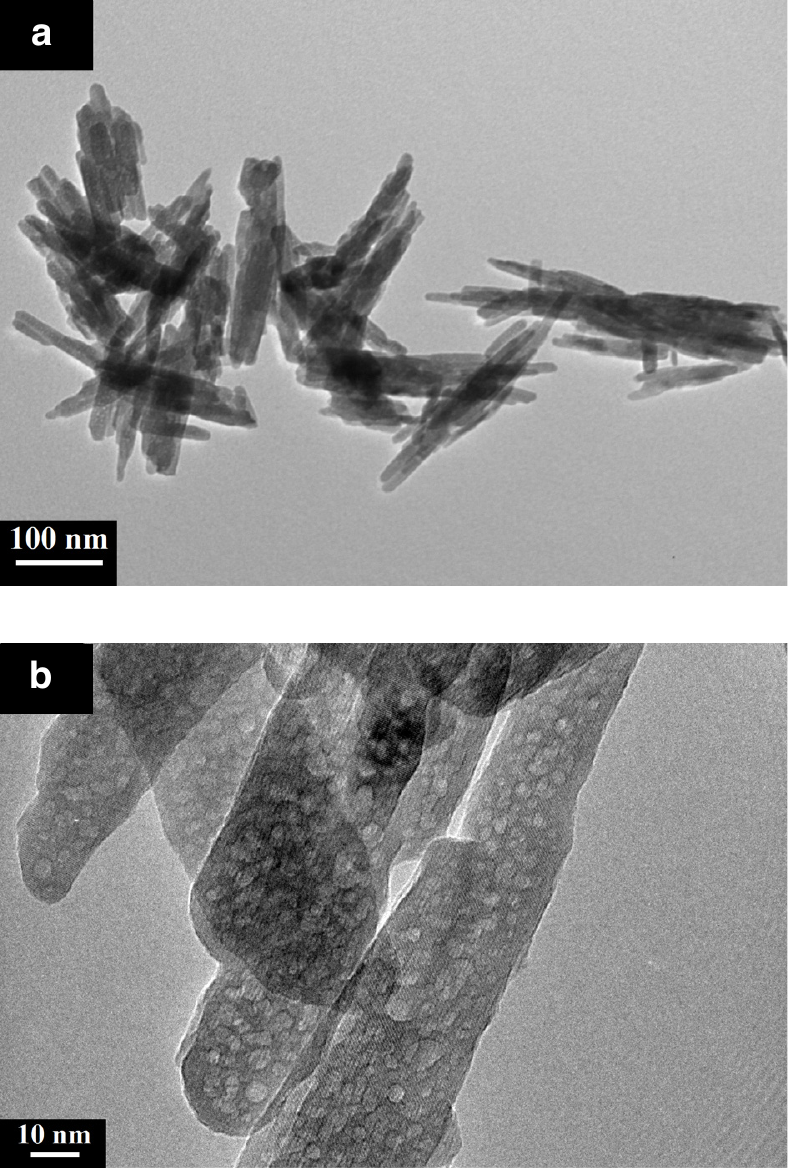}
  \caption{TEM images of hydroxyapatite particles at two different magnifications: (a) $\times150,\!000$ and (b) $\times10^6$.}
  \label{Fig:4}
\end{figure}

Titanium foil (0.127~mm), \ce{TiO2} (rutile, the particle size less than 100~nm), poly(styrene-alt-maleic acid), \ce{Ca(NO3)2*4H2O}, \ce{(NH4)2HPO4}, \ce{NH4OH}, and Hank's balanced salt solution (\ce{CaCl2} 0.14~g/L, KCl 0.40~g/L, \ce{KH2PO4} 0.06~g/L, \ce{MgCl2*6H2O} 0.10~g/L, \ce{MgSO4*7H2O} 0.10~g/L, NaCl 8.00~g/L, \ce{NaHCO3} 0.35~g/L, \ce{Na2HPO4} 0.048~g/L, glucose 1.00~g/L, phenol red 0.01~g/L, Sigma-Aldrich, Canada) were used for the following experiments. Stoichiometric HA was synthesized using wet chemical precipitation, by the slow addition of 0.6~M \ce{(NH4)2HPO4} solution to 1.0~M \ce{Ca(NO3)2} solution while continuously stirring at 70$^\circ$C. The solution was stirred for 8~h at 70$^\circ$C, followed by 24~h of stirring at room temperature, and the pH was adjusted to 11 using \ce{NH4OH}. The resulting precipitate was washed using water and ethanol, then dried. Crystalline needle-shaped HA nanoparticles have been obtained using this method. A TEM micrograph presented in Fig.~\ref{Fig:4} shows  a needle-shaped HA particle morphology. HA particles had an aspect ratio of approximately eight, and an average length of 150~nm.

Stainless steel or titanium foil were used as substrates for EPD. The substrate was placed 15~mm away from the platinum counter electrode.  The deposition time was varied in the range of 1$-$6 mins, and deposition voltage was within the range of 10$-$15~V. \ce{TiO2} and HA were dispersed in a mixed water-ethanol solvent (40\% water), containing dissolved PSMA-h. The use of the mixed solvent offered the advantages of reduced gas evolution at the substrate surface. The deposition yield was studied using stainless steel substrates. Potentiodynamic studies were performed using  a Parstat 2273 potentiostat and the PowerSuite software (Princeton Applied Research). A three-cell electrode cell was used for electrochemical testing, with a saturated calomel electrode (SCE) as the reference electrode and platinum mesh as the counter electrode. The test was carried out in Hank's balanced salt solution, which was used to simulate physiological conditions. A scan rate of 1~mV/s was used to obtain the potentiodynamic polarization curves.

A JEOL 7000F scanning electron microscope (SEM) and FEI Tecnai Osiris transmission electron microscope (TEM) were used for electron microscopy. A Nicolet I2 diffractometer with monochromatized CuK$_\alpha$ radiation was used for X-Ray diffraction (XRD). Fourier-transform infrared spectroscopy (FTIR) studies were performed on Bruker Vertex 70 spectrometer.

%%%%%%%%%%%%%%%%%%%%%%%%%%%%%%%%%%%%%%%%%%%%%%%%%%%%%%%%%%%%%%%%%%%%%%%%%%%%%%%%%%%%%%%%%%%
\section{Results and discussion}

We begin with examining surfaces of rutile and HA. The specific surface energy is defined as
\begin{equation}\label{Eq:1}
  \gamma = \frac{E_\text{tot}(\text{slab})-E_\text{tot}(\text{bulk})}{2A},
\end{equation}
where $E_\text{tot}$  is the DFT total energy of the slab and the bulk material, $A$ is the planar surface area, and the factor of 2 accounts for the present of two surfaces at the top and bottom of the simulation slab. Results for the surface energy are presented in Table~\ref{Tab:2}, which also includes values of the surface energies reported in the literature for comparison. A large scattering of literature values, in particular for HA, can be attributed to variability in DFT approximations for the exchange-correlation functional, selection of pseudopotentials, and uncertainties in the structure specifically for the HA$(01\overline{1}0)$ surface. HA has a higher surface energy that correlates with its higher wettability in comparison to \ce{TiO2}. The presence of a polar solvent (water) significantly affects the surface energy, in particular for HA. The disparity between the surface energies of \ce{TiO2} and HA vanishes in the solvent.

\begin{table}
  \caption{\ Calculated surface energy and absorption energy of MA at \ce{TiO2} and HA surfaces. Values are presented for various solvent conditions: vacuum/mixture 80\% ethanol + 20\% water/100\% water, respectively.}
  \label{Tab:2}
  \begin{tabular}{lccc}
    \hline
    Surface & Surface energy & \multicolumn{2}{c}{Adsorption enthalpy (eV)} \\
     & (J/m$^2$) & monomer & segment \\
    \hline
    \ce{TiO2} $(110)$ & 0.63\textsuperscript{\emph{a}}/0.48/0.44 & $-$1.3/$-$0.7/$-$0.6 & $-$1.3/$-$0.9/$-$0.6 \\
    HA $(0001)$ & 0.90\textsuperscript{\emph{b}}/0.53/0.45 & $-$3.4/$-$1.3/$-$1.2 & --- \\
    HA $(01\overline{1}0)$ & 0.96$^c$/0.59/0.51 & $-$4.7/$-$2.5/$-$2.1 & $-$4.0/$-$1.8/$-$1.5 \\
    \hline
  \end{tabular}
  
  \textsuperscript{\emph{a}} 0.31$-$0.47, 0.89 J/m$^2$ (DFT calculations\cite{Bourikas_CR_114_2014,Ramamoorthy_PRB_49_1994});
  \textsuperscript{\emph{b}}~0.33, 0.77, 1.01, 1.93 J/m$^2$ (DFT calculations\cite{Zhu_CPL_396_2004,Chiatti_TCA_135_2016,Almora-Barrios_L_26_2010,Zhao_L_30_2014});
  \textsuperscript{\emph{c}}~0.33, 1.32, 1.36, 2.10 J/m$^2$ (DFT calculations\cite{Zhu_CPL_396_2004,Almora-Barrios_L_26_2010,Chiatti_TCA_135_2016,Zhao_L_30_2014})
\end{table}

MA molecule and MA segment adapt a three-dimensional structure as shown in Figs.~\ref{Fig:3}a,b and \ref{Fig:3}e,f, respectively. Some adsorption configurations require a structural deformation (flattening) of the MA molecule or MA segment. The flattening occurs \textit{via} rotating the carboxyl groups to align both groups to a common plane as illustrated in Figs.~\ref{Fig:3}c,d and \ref{Fig:3}g,h, respectively. The associated deformation energy is of the order of 0.15~eV as evaluated by DFT calculations, which is one order of magnitude less than typical adsorption energies. The ability of MA to easily adapt its structure to the surface of interest can be attributed to its aliphalic nature.

\begin{figure}
 \centering
 \includegraphics{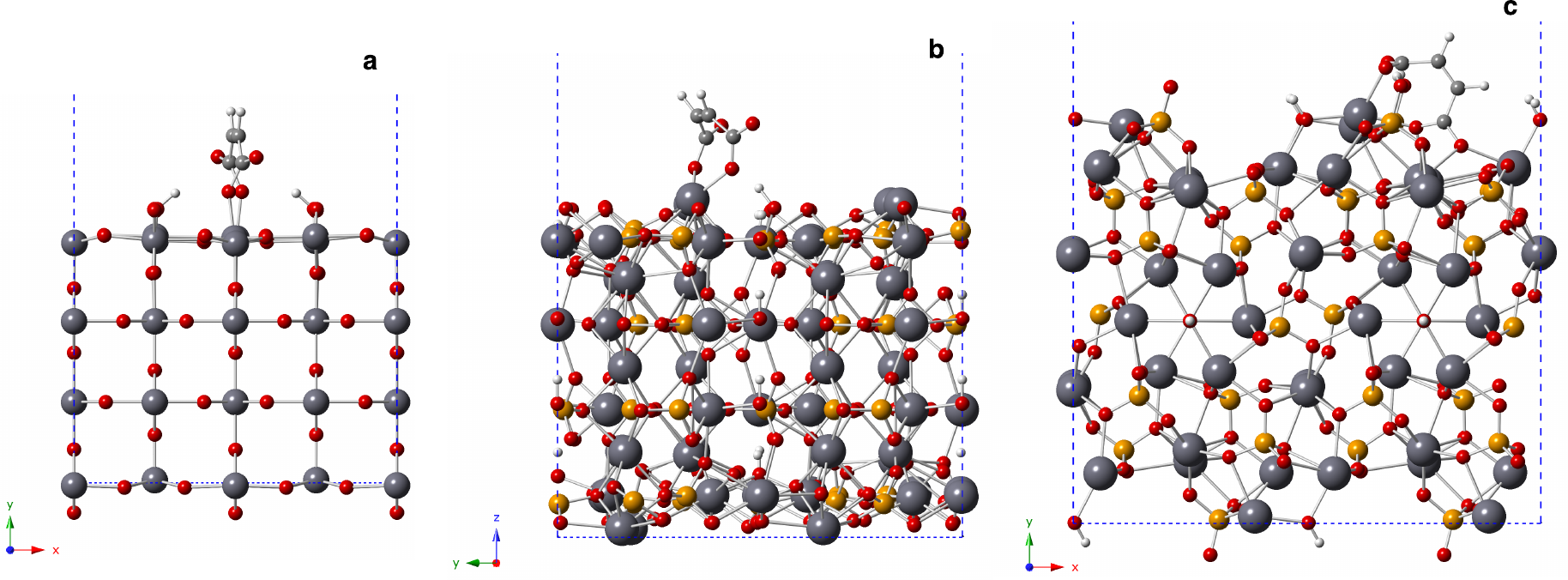}
 \caption{(a) Maleic acid monomer at the $(110)$ surface of \ce{TiO2}, (b) at the HA $(0001)$ surface and (c) at the HA $(01\overline{1}0)$ surface.}
 \label{Fig:5}
\end{figure}

\begin{figure}%[h]
\centering
  \includegraphics{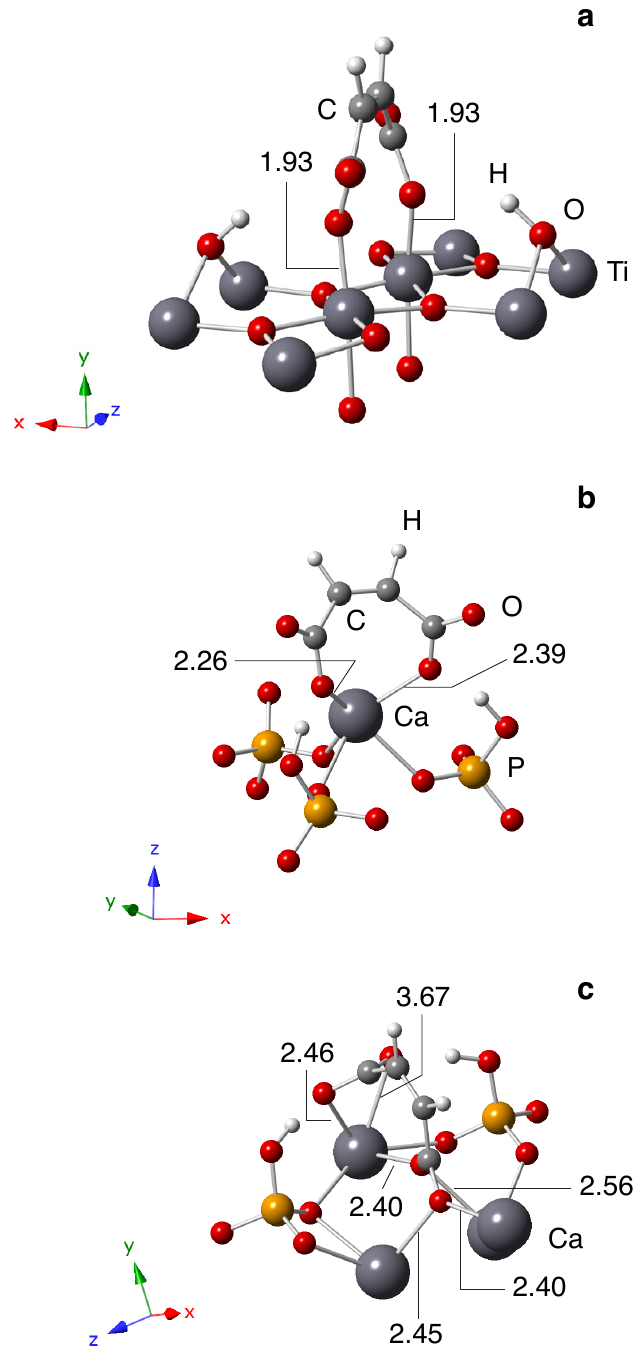}
  \caption{Maleic acid and neighbour atoms of the adsorption complex (a) at the $(110)$ surface of \ce{TiO2}, (b) at the HA $(0001)$ surface, and (c) at the HA $(01\overline{1}0)$ surface. Bond distances between the maleic acid and surface metal atoms are shown in Angstroms for an aqueous environment.}
  \label{Fig:6}
\end{figure}

Figure~\ref{Fig:5} shows the lowest energy configuration of MA monomer adsorbed on \ce{TiO2} and HA surfaces. Adsorption to the \ce{TiO2} surface takes place \textit{via} an inner sphere bonding to a pair of Ti-atoms (a bridge bidentate coordination) as illustrated in Fig.~\ref{Fig:6}a, rather than chelate bidentate bonding to a single Ti site at the surface. This result can be explained by a dense packing of Ti atoms at \ce{TiO2} surface and is reminiscent of catecholates' adsorption on \ce{TiO2}.\cite{Ye_CSR_40_2011,Zhang_RA_5_2015} In the case of HA, the spacing between Ca-atoms at the surface is too large. Therefore, chelation is a preferable type of bonding on HA (Fig.~\ref{Fig:6}b,c). The adsorption of MA is accompanied by deprotonation of both carboxyl groups. Those protons are readily attracted to passivate oxygen dangling bonds at the surface \ce{PO4}-tetrahedra and to restore the change balance perturbed by the newly formed \ce{Ti-O} or \ce{Ca-O} bonds.\cite{Zhang_RA_5_2015} Regarding \ce{H+} placement on the HA$(01\overline{1}0)$ surface, there are several alternatives to the top of \ce{PO4}-tetrahedra, which include O-atoms in the base of \ce{PO4}-tetrahedra as well as the \ce{\bond{#}Ca2OH^{$q$-}} groups present at the surface. It turns out that \ce{\bond{#}Ca2OH2^{1-$q$}} is the second most favourable scenario from DFT the total energy point of view.

The affinity of MA to the surfaces is characterized by the adsorption enthalpy, which is defined as
\begin{equation}\label{Eq:2}
  \Delta H_\text{ads} \approx E_\text{tot}(\text{MA~ads.})-E_\text{tot}(\text{MA})-E_\text{tot}(\text{slab}),
\end{equation}
where $E_\text{tot}(\text{MA~ads.})$ is the total energy of a slab with the molecule adsorbed on its surface (Fig.~\ref{Fig:5}), $E_\text{tot}(\text{MA})$ is the total energy of the molecule or segment in its ground state configuration (Figs.~\ref{Fig:3}a or \ref{Fig:3}e), and $E_\text{tot}(\text{slab})$ is the energy of the slab. The approximate sign reflects neglecting a zero-point energy change upon adsorption and finite temperature effects. This approximation is justified in an earlier study of catalytic reactions,\cite{Gross_JVSTA_15_1997} since the sum of zero-point energies is approximately constant during the traversing of regions with energy barriers.

The adsorption enthalpies of MA on \ce{TiO2} and HA surfaces are summarized in Table~\ref{Tab:2}. Here we list results obtained for different solvents: water, a mixture of water and ethanol as well as for the vacuum since effects of the solvent are often neglected in studies of a molecular adsorption. Values of the adsorption enthalpy in Table~\ref{Tab:2} suggest that the binding energies of molecules to the surface are significantly overestimated in vacuum (almost by a factor of two). The more ionic the solvent, the weaker is the affinity of MA to the surface. The adsorption enthalpies of MA on \ce{TiO2} and HA surfaces in water remain strong: $\Delta H_\text{ads}=-0.6$ and $-1.5$~eV respectively, in spite of solvation effect. The values are comparable to $-1$~eV bond strength between DOPA and Si surface measured experimentally,\cite{Lee_PNASU_103_2006} the theoretical range from $-$1.1 to $-$1.3~eV for the \ce{TiO2}-dopamine bond strength,\cite{Vega-Arroyo_CPL_406_2005} and $-1.5$~eV for the caffeic acid adsorption on \ce{TiO2} in vacuum.

Effect of the solution pH on adsorption energies can be accounted by evaluating a change in the free energy of a residue as a result of deprotonation\cite{Bombarda_JPCB_114_2010}
\begin{equation}\label{Eq:3}
  \Delta G=-RT\ln 10(\mathrm{pH}-\mathrm{p}K_\text{a}),
\end{equation}
where $R$ is the gas constant, $T$ is the temperature, and $\mathrm{p}K_\text{a}$ is a dissociation constant. The dissociation constants for MA are $\mathrm{p}K_{\text{a}1}=1.9$ and $\mathrm{p}K_{\text{a}2}=6.0$\cite{Pirrone_JBB_2010_2010}. This implies that both carboxyl groups are deprotonated at the experimental conditions ($\mathrm{pH}=7$). As a result, the free energy of MA is lowered by $\Delta G\approx-0.16$~eV per residue ($-0.10$ and $-0.06$~eV for the first and second group, respectively). To account for deprotonation, the magnitude of adsorption enthalpies in Table~\ref{Tab:2} should be corrected by the corresponding value of $\Delta G$, which results in reduction of the magnitude of the adsorption energy with increasing pH value but does not prevent MA residues from adsorption as observed experimentally by \citet{Hidber_JECS_17_1997}.

\begin{figure}%[h]
\centering
  \includegraphics{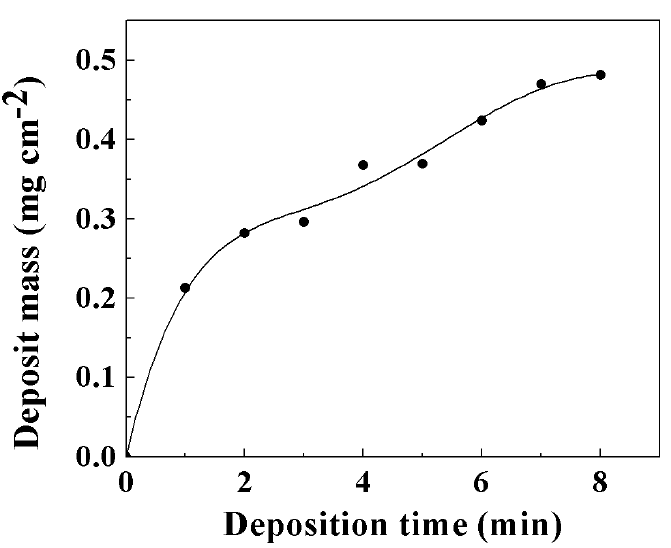}
  \caption{Deposit mass versus deposition time for 5~g~L$^{-1}$ \ce{TiO2} suspension, containing 10~g~L$^{-1}$ PSMA-h at a deposition voltage of 15~V.}
  \label{Fig:7}
\end{figure}

Next we continue with an experimental verification of theoretical predictions. It should be emphasized once again that individual MA monomers are used in the modeling section for the sake of computational simplicity, while PSMA-h copolymers are employed in the experimental section to provide multiple chemical bonds with substrates.  Due to the limitations of the concept of zeta-potential for analysis of particles, containing adsorbed polyelectrolytes,\cite{Zhitomirsky_ACIS_97_2002,Ohshima_CSA_103_1995,Ohshima_JCIS_233_2001} the influence of PSMA-h adsorption on the electrokinetic behavior of the \ce{TiO2} (rutile) particles was analyzed using the EPD yield data. The EPD experiments performed with the 5$-$10~g~L$^{-1}$ \ce{TiO2} (rutile) suspensions without PSMA-h showed the formation of cathodic deposits. The low cathodic EPD yield indicated that particles were weakly positively charged. PSMA-h films were deposited anodically and showed strong adhesion to the substrates. The film adhesion corresponded to  5B classification (ASTM~D3359). The addition of PSMA-h to the \ce{TiO2} suspensions resulted in significant improvement of the suspension stability and formation of anodic deposits.  The formation of anodic films signified a particle charge reversal due to the adsorption of the negatively charged PSMA-h. The deposit mass increased with increasing deposition time at a constant applied voltage (Fig.~\ref{Fig:7}). The experimental results indicated that the film thickness can be controlled and varied. The deposition yield data showed relatively high deposition rate. A decrease of deposition rate with time was observed due to the reduction in the voltage drop in the bulk of the suspension as a result of the formation of an insulating film layer on the electrode surface.

\begin{figure}%[h]
\centering
  \includegraphics{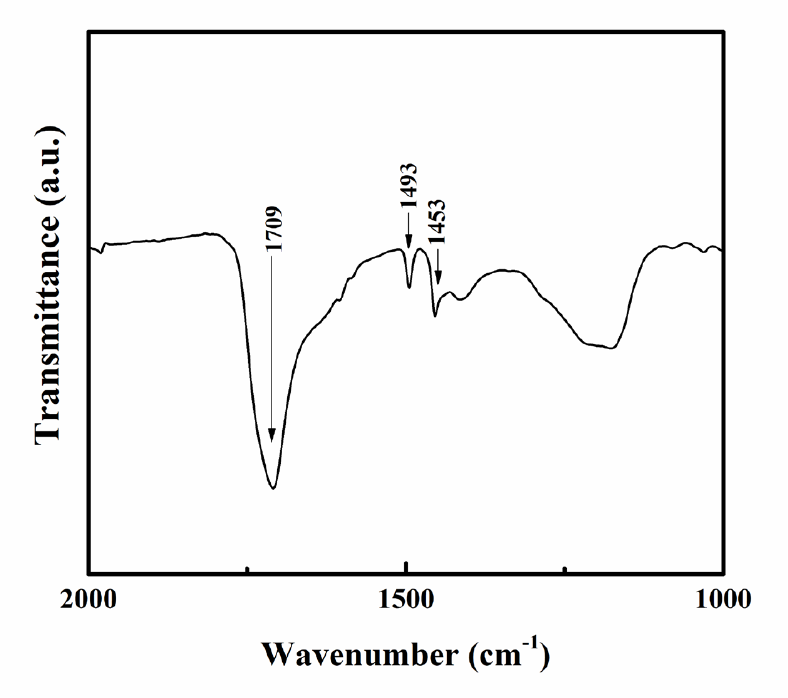}
  \caption{FTIR spectrum of a deposit, obtained from a suspension, containing 10~g~L$^{-1}$ \ce{TiO2} and 10~g~L$^{-1}$ PSMA-h.}
  \label{Fig:8}
\end{figure}

The PSMA-h adsorption on the \ce{TiO2} particles was confirmed by the FTIR analysis (Fig.~\ref{Fig:8}) of the deposited  material. Figure~\ref{Fig:8} shows absorption peaks at 1453, 1493~cm$^{-1}$ related to \ce{C-C} vibrations of the aromatic rings of the styrene monomers and another absorption at  1709 cm$^{-1}$ attributed to \ce{C=O} vibrations of the MA monomers of the adsorbed polymer.

\begin{figure}%[t]
\centering
  \includegraphics{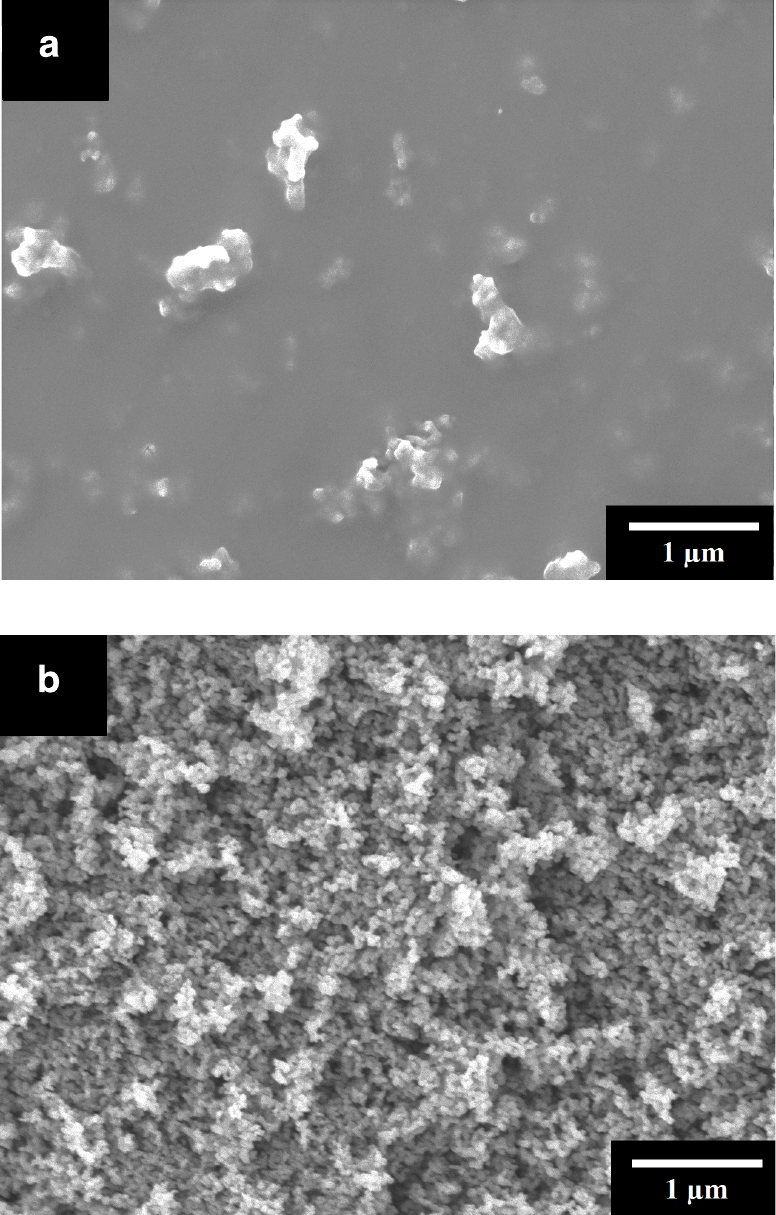}
  \caption{SEM images of deposits, prepared from (a) 5 and (b) 10~g~L$^{-1}$ \ce{TiO2} suspension, containing 10~g~L$^{-1}$ PSMA-h at a deposition voltage of 10~V.}
  \label{Fig:9}
\end{figure}

SEM studies showed the formation of continuous and crack-free \ce{TiO2}-PSMA-h films (Fig.~\ref{Fig:9}). It was found that the variation of \ce{TiO2} concentration in the suspension resulted in changes of film microstructure. The SEM image of a film, prepared from the 5~g~L$^{-1}$ \ce{TiO2} suspension, containing 10~g~L$^{-1}$ PSMA-h showed \ce{TiO2} particles in a PSMA-h matrix (Fig.~\ref{Fig:9}a).  The SEM image of the film prepared from 10~g~L$^{-1}$ \ce{TiO2} suspension (Fig.~\ref{Fig:9}b), containing 10~g~L$^{-1}$ PSMA-h showed mainly \ce{TiO2} particles, which formed a porous film. The film porosity resulted from packing of \ce{TiO2} particles. The comparison of the SEM images, shown in Fig.~\ref{Fig:9}a,b, indicated that the increase of \ce{TiO2} concentration in the 10~g~L$^{-1}$ PSMA-h solution resulted in the increasing \ce{TiO2} content in the deposited film.

\begin{figure}%[h]
\centering
  \includegraphics{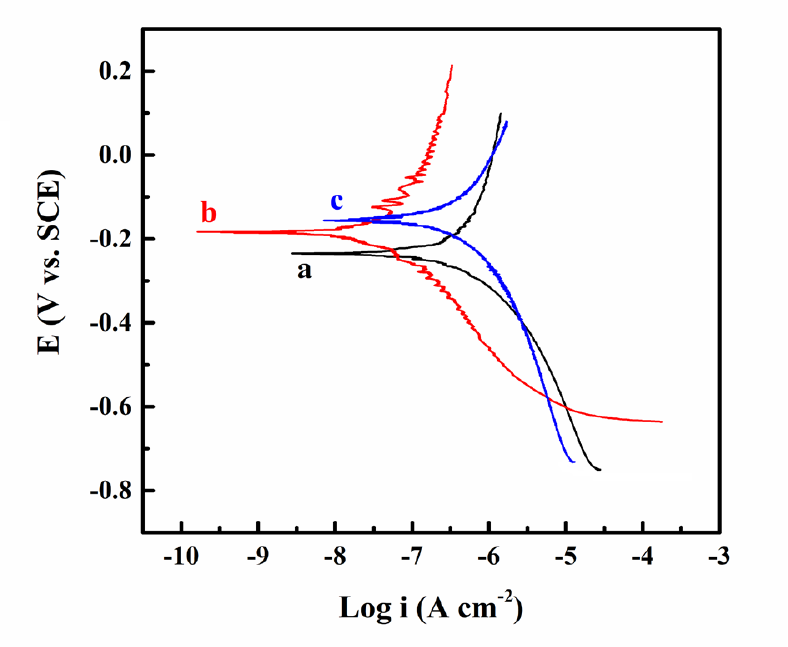}
  \caption{Tafel plots in Hank's solutions for (a) uncoated Ti,  (b) coated by deposition from 5~g~L$^{-1}$ \ce{TiO2} suspension, containing 10~g~L$^{-1}$ PSMA-h and (c) coated by deposition from a suspension, containing 5~g~L$^{-1}$ \ce{TiO2} , 5~g~L$^{-1}$ HA and  10~g~L$^{-1}$ PSMA-h at a deposition voltage of 10~V during 5~min.}
  \label{Fig:10}
\end{figure}

The composite \ce{TiO2}-PSMA-h coating prepared from 5~g~L$^{-1}$ \ce{TiO2} suspension, containing 10~g~L$^{-1}$ PSMA-h, was studied in Hank's balanced salt solution, which acted as a simulated body fluid. Tafel plots comparing the electrochemical behaviour of the coated and uncoated titanium are shown in Fig.~\ref{Fig:10}. From the Tafel plots it can be seen that deposited \ce{TiO2}-PSMA-h coating allowed a reduction of the corrosion current, compared to an uncoated titanium. Moreover, the coated substrate showed a higher corrosion potential. These results demonstrated that the coating acted as a protective layer and provided corrosion protection of the titanium substrates. Therefore the \ce{TiO2}-PSMA-h coating, containing the bioactive \ce{TiO2} rutile phase offers additional benefits of corrosion protection of underlying metallic substrates for biomedical implant applications.

\begin{figure}[t]
\centering
  \includegraphics{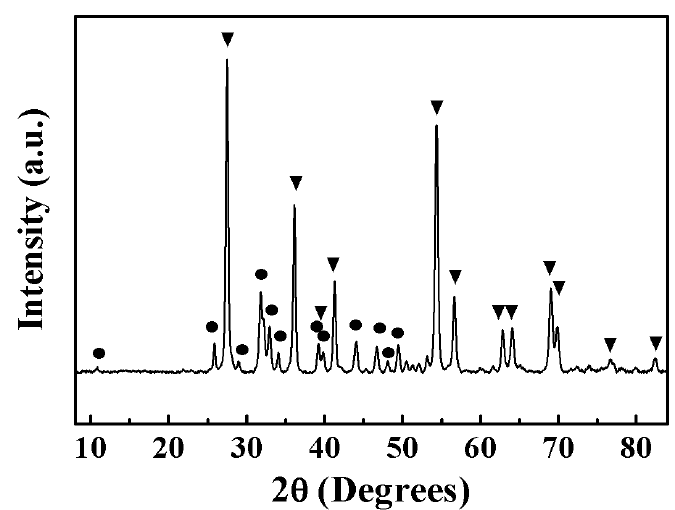}
  \caption{X-ray diffraction pattern of a composite coating, prepared from a suspension, containing 5~g~L$^{-1}$ HA and 5~g~L$^{-1}$ \ce{TiO2}, containing 10~g~L$^{-1}$ PSMA-h at a deposition voltage of 10~V ($\blacktriangledown$-\ce{TiO2} rutile, $\medblackcircle$-HA).}
  \label{Fig:11}
\end{figure}

\begin{figure}%[t]
\centering
  \includegraphics{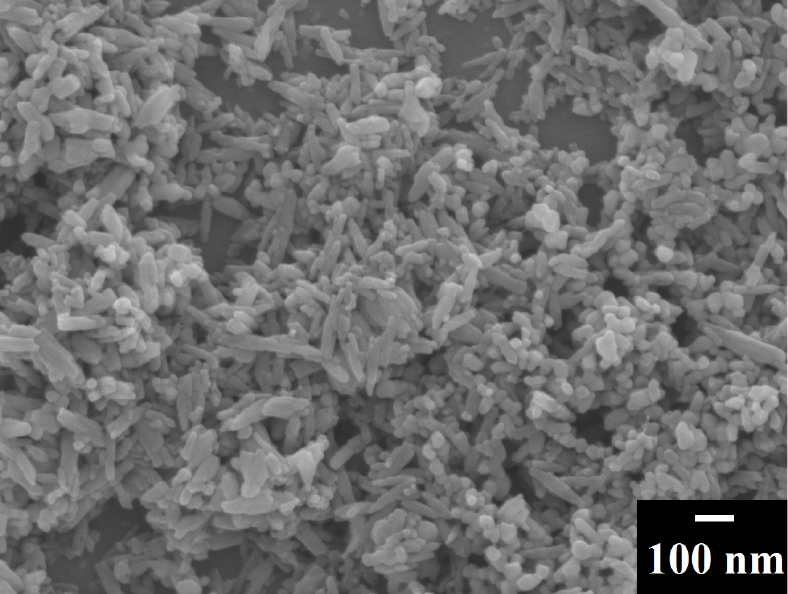}
  \caption{SEM image of the composite film, prepared from the suspensions, containing 5~g~L$^{-1}$ \ce{TiO2}, 5~g~L$^{-1}$ HA and 10~g~L$^{-1}$ PSMA-h.}
  \label{Fig:12}
\end{figure}

Related to biomedical applications, we investigated PSMA-h as a co-dispersant for the co-deposition of \ce{TiO2} (rutile) and HA to fabricate composite rutile-HA coatings. A suspension containing 5~g~L$^{-1}$ HA, 5~g~L$^{-1}$ \ce{TiO2} and 10~g~L$^{-1}$ PSMA-h was used for anodic EPD, and subsequent films were studied using XRD. The resulting X-ray diffraction pattern is shown in Fig.~\ref{Fig:11}. The XRD pattern showed \ce{TiO2} (rutile) peaks, corresponding to the JCPDS file 021-1276 and peaks of HA, corresponding to the JCPDS file 046-0905. This confirmed that HA and \ce{TiO2} were co-deposited using PSMA-h, and thus composite HA-rutile-PSMA-h films were formed. The results of SEM studies (Fig.~\ref{Fig:12}) provided additional evidence of the formation of composite coatings. The SEM image presented in Fig.~\ref{Fig:12} shows needle-shape HA particles in addition to the \ce{TiO2} particles. Therefore, PSMA-h can be used as a co-dispersing and film-forming agent for the co-deposition of HA and rutile. The composite coatings showed corrosion protection of Ti substrates, as indicated (Fig.~\ref{Fig:10}) by the increase in the corrosion potential and reduction of the corrosion current. As emphasised above, the composite coating containing HA and rutile offer many benefits for biomedical application. EPD method has  many processing advantages, such as high deposition rate and possibility of uniform deposition on substrates of complex shape and high surface area.

Finally, we would like to comment on an interference between adsorption of MA and hydration of surface by chemisorption of water molecules that takes place under relevant experimental conditions. The hydration of metal oxides is governed by the following two-step reaction 
\begin{subequations}\label{Eq:4}
	\begin{align}
		\ce{\bond{#}Me^{$q$+} + H2O &<=> \bond{#}MeOH^{$q$-1} + H^+(aq)}\label{Eq:4a}\\
		\ce{\bond{#}O^{$q$-} + H^+(aq) &<=> \bond{#}OH^{1-$q$}}\label{Eq:4b}
	\end{align}
\end{subequations}
Here \ce{\bond{#}Me} is a surface metal ion (either Ti or Ca), and $q$ indicates its fractional charge. The charge is $q=2/3e$ for rutile $(110)$ surface,\cite{Zhang_RA_5_2015} where $e$ is the elementary charge.  The HA$(01\overline{1}0)$ surface has several \ce{\bond{#}Ca^{$q$+}} and \ce{\bond{#}O^{$q$-}} sites with a range of $q$-values up to $1e$ due to a variety of dangling bonds created in the process of cleaving the surface (see Fig.~\ref{Fig:2}). Results for the enthalpy of water molecule chemisorption at these two surfaces are listed in Table~\ref{Tab:3}. The water chemisorption is favourable at both surfaces in vacuum, which agrees with previous studies.\cite{Zheng_JCP_145_2016,Chiatti_TCA_135_2016} However, the presence of solvent not only reduces the magnitude of interaction but also changes the surface receptivity. As evident from Table~\ref{Tab:3}, the enthalpy of dissociative adsorption of water molecules at the HA$(01\overline{1}0)$ surface in aqueous environment is positive, indicating that its chemisorption is unlikely. When reflected onto experimental conditions, this result implies that the surface of rutile particles is likely to be terminated with hydroxyl groups during a reaction with PSMA-h, whereas the surface of HA is not. As a benchmark for the solvation model we also computed the vaporization enthalpy of water molecule $+$0.32~eV \textit{vs} the experimental value of $+$0.46~eV,\cite{cox1989codata} which sets error-bars of the calculation.

Adsorption of MA on the rutile $(110)$ surface in the presence of surface hydroxyl groups is described as a multistep ligand (L) exchange reaction\cite{Hidber_JECS_17_1997,Rodriguez_JCIS_177_1996}
\begin{subequations}\label{Eq:5}
	\begin{align}
		\ce{LH2 (aq) &<=> L^{2-}(aq) + 2H+(aq)}\label{Eq:5a}\\
		\ce{2\bond{#}TiOH^{1/3-} + 2H+(aq) &<=> 2\bond{#}TiOH2^{2/3+}}\label{Eq:5b}\\
		\ce{2\bond{#}TiOH2^{2/3+} + L^{2-}(aq) &<=> \bond{#}Ti2L^{2/3-} + 2H2O}\label{Eq:5c}
	\end{align}
\end{subequations}
which involves deprotonation of the ligand, protonation of surface hydroxyl groups on Ti-site, and finally the ligand exchange. This reaction leads to desorption of two water molecules from the surface of \ce{TiO2}, which is energetically unlikely in an aqueous environment due to reasons discussed in the preceding paragraph. As a result, an enthalpy is positive for the ligand  exchange reaction between hydroxyl groups at the \ce{TiO2} surface and MA monomer or MA segment as a part of PSMA-h (Table~\ref{Tab:3}). To explain the experimentally observed adsorption of PSMA-h on \ce{TiO2}, we recall that the adsorption takes place in in a water-ethanol solution. In their experimental study of adsorption of alcohols on \ce{TiO2} (rutile) surface, \citet{Suda_L_3_1987} noted that ethanol adsorption to the surface results in expelling water from the surface, which implies presence of \ce{\bond{#}Ti^{2/3+}} bonding site at the surface with no need for the ligand exchange.

The absence of a strong dissociative bond between water and HA allows us to assume coexistence of the hydrated \ce{\bond{#}CaOH} and \ce{\bond{#}OH} surface sites with unhydrated \ce{\bond{#}Ca^{$q$+}} and \ce{\bond{#}O^{$q$-}} sites. The ligand exchange reaction in this case proceeds through the following steps
\begin{subequations}\label{Eq:6}
	\begin{align}
		\ce{\bond{#}CaOH + \bond{#}O^{$q$-} + 2H+(aq) &<=> \bond{#}CaOH2+ + \bond{#}OH^{1-$q$}}\label{Eq:6a}\\
		\ce{\bond{#}CaOH2^{+} + \bond{#}Ca^{$q$+} + L^{2-}(aq) &<=> \bond{#}Ca2L^{$q$-1} + H2O}\label{Eq:6b}
	\end{align}
\end{subequations}
The corresponding enthalpy is strongly negative (see Table~\ref{Tab:3}) indicating that, unlike in the case of \ce{TiO2}, there is no competition between water chemisorption and MA adsorption at the surface of HA.

\begin{table}
  \caption{\ Chemisorption enthalpy of water and MA molecules at rutile $(110)$ and HA$(01\overline{1}0)$ surfaces calculated in a gas phase and in an aqueous solution.}
  \label{Tab:3}
  \begin{tabular}{lccc}
    \hline
    Species & Reactions & \multicolumn{2}{c}{Enthalpy (eV)} \\
     &  & in gas phase & aq. solution \\
    \hline
    \ce{H2O} on \ce{TiO2} & Eq.~(\ref{Eq:4}) & $-$0.9 & $-$0.4 \\
    \ce{H2O} on HA & Eq.~(\ref{Eq:4}) & $-$1.8 & $+$0.3 \\
    MA monomer on \ce{TiO2} & Eq.~(\ref{Eq:5}) & --- & $+$0.3 \\
    MA segment on \ce{TiO2} & Eq.~(\ref{Eq:5}) & --- & $+$0.4 \\
    MA monomer on HA & Eqs.(\ref{Eq:5a}) and (\ref{Eq:6}) & --- & $-$2.2 \\
    MA segment on HA & Eqs.(\ref{Eq:5a}) and (\ref{Eq:6}) & --- & $-$1.6 \\
    \hline
  \end{tabular}
\end{table}

%%%%%%%%%%%%%%%%%%%%%%%%%%%%%%%%%%%%%%%%%%%%%%%%%%%%%%%%%%%%%%%%%%%%%%%%%%%%%%%
\section{Conclusions}

Poly(styrene-alt-maleic acid) adsorption on hydroxyapatite and \ce{TiO2} (rutile) was studied using
experimental techniques and corroborated by \textit{ab initio} simulations of adsorption of a maleic acid segment as a subunit of the copolymer. \textit{Ab initio} calculations suggest that the maleic acid segment adsorbs to the \ce{TiO2} surface \textit{via} an inner sphere bridge bidentate bonding to a pair of Ti-atoms. Chelation is a preferable type of bonding for the maleic acid segment on hydroxyapatite due to peculiarities of the surface reconstruction. The aliphatic nature of maleic acid makes it adaptive to various surfaces with a little energy penalty (of the order 0.15~eV) associated with structural changes such as flattening. A magnitude of the adsorption enthalpy for the maleic acid monomer and the segment is comparable to that of catecholates. Solvent effects play a twofold role when providing quantitative evaluation of the adsorption energies and, thus, cannot be neglected. First, it significantly reduces the adsorption strength (by almost factor of two if compared to vacuum) as the polarity of the solvent increases. Second, a water chemisorption at the surface hinders the adsorption of maleic acid at the surface of rutile. The results of first-principle calculations were confirmed by the experimental measurements. We found that adsorbed poly(styrene-alt-maleic acid) allowed efficient dispersion of rutile and formation of films by the electrophoretic deposition. We investigated the deposition yield and morphology of the films. Moreover, it was found that rutile can be co-dispersed and co-deposited with hydroxyapatite to form composite films. The coatings obtained by the electrophoretic deposition showed corrosion protection of metallic implants in simulated body fluid solutions, which is a favorable characteristic for biomedical applications.

%%%%%%%%%%%%%%%%%%%%%%%%%%%%%%%%%%%%%%%%%%%%%%%%%%%%%%%%%%%%%%%%%%%%%
\begin{acknowledgement}

M.A. and O.R. would like to acknowledge the funding provided by the Natural Sciences and Engineering Research Council of Canada under the Discovery Grant Program RGPIN-2015-04518. The computations were performed using Compute Canada (Calcul Quebec and Compute Ontario) resources, including the infrastructure funded by the Canada Foundation for Innovation. A.C. and I.Zh. would like to acknowledge the funding provided by the Natural Sciences and Engineering Research Council of Canada under the Strategic Project 447475-13.

\end{acknowledgement}

%%%%%%%%%%%%%%%%%%%%%%%%%%%%%%%%%%%%%%%%%%%%%%%%%%%%%%%%%%%%%%%%%%%%%
\section*{Notes}
There are no conflicts to declare.

%\bibliography{/Users/oleg/Documents/SHARED-DOCS/PAPERS/bibliography}

\providecommand{\latin}[1]{#1}
\makeatletter
\providecommand{\doi}
  {\begingroup\let\do\@makeother\dospecials
  \catcode`\{=1 \catcode`\}=2 \doi@aux}
\providecommand{\doi@aux}[1]{\endgroup\texttt{#1}}
\makeatother
\providecommand*\mcitethebibliography{\thebibliography}
\csname @ifundefined\endcsname{endmcitethebibliography}
  {\let\endmcitethebibliography\endthebibliography}{}

\end{document}